\documentclass[prl,twocolumn,showpacs,amsmath,amssymb]{revtex4}%
\usepackage[dvips]{graphicx}
\usepackage{dcolumn}
\usepackage{bm}

\begin{document}

\title{Evidence for Two-Dimensional Spin-Glass Ordering in Submonolayer Fe Films\\
on Cleaved InAs Surfaces}

\author{Toshimitsu Mochizuki}
\author{Ryuichi Masutomi}
\author{Tohru Okamoto}
\affiliation{%
Department of Physics, University of Tokyo,
7-3-1, Hongo, Bunkyo-ku, Tokyo 113-0033, Japan}%

\date{October 24, 2008}

\begin{abstract}
Magnetotransport measurements have been performed on
two-dimensional electron gases formed at InAs(110) surfaces
covered with a submonolayer of Fe.
Hysteresis in the magnetoresistance, 
a difference in remanent magnetoresistance
between zero-field-cooling procedures and field-cooling procedures,
and logarithmic time-dependent relaxation after magnetic field sweep 
are clearly observed at 1.7~K for a coverage of 0.42~monolayer.
These features are associated with spin-glass ordering in the Fe film.
\end{abstract}

\maketitle

Spin glasses are magnetic systems 
with randomly competing interactions.
They have attracted great interest
during the last few decades \cite{EA, SK, Youngreview}.
Spin-glass models and related methods have also been useful
in other areas of science such as simulation of protein folding \cite{Bryngelson}
and optimization problems in computer science \cite{optimization}.
Most of the attempts to understand spin glasses have been
concerned with the behavior in three dimensions.
It is generally believed that, in two dimensions, the spin-glass ordering does not occur 
at nonzero temperature.
The lower critical dimension of spin-glass ordering has been shown to be $d_l > 2$
for Heisenberg spins \cite{Banavar1982} and $XY$ spins \cite{Maucourt1998}, and Ising spins with Gaussian distribution of disorder \cite{McMillan1984,Bray1984}.
Although the situation for the Ising model with bimodal ($\pm J$) disorder was controversial \cite{Shirakura, Matsubara},
recent theoretical investigations do not support the existence of the spin-glass phase for $T>0$ \cite{Kawashima1997,HartmannYoung2001,Houdayer2001}.
On the other hand, numerical calculations have demonstrated that 
the spin-glass-like ordering temperature can be nonzero
for a two-dimensional (2D) Ising model with random nearest-neighbor interactions and
ferromagnetic second-neighbor interactions \cite{Lemke1996,HartmannCampbell2001}.
Experimentally, spin-glass behavior was found 
in thin films \cite{Sandlund,Hoines,Granberg,Morenzoni}
and layered compounds \cite{Gavrin1991,Mori2003,Mathieu2007}.
However, no observation has been reported
for a single layer system with strict two dimensionality.

Submonolayer films of magnetic materials adsorbed on nonmagnetic substrates 
are promising candidates for 2D spin-glass systems.
A random distribution of adsorbates can be obtained
by deposition at low substrate temperatures,
where surface diffusion is minimal and island growth is limited.
In the case that the sign of the interaction depends 
on the relative position of adatoms,
a competition between ferromagnetic and antiferromagnetic interactions is expected.
As the substrate, narrow band-gap III-V semiconductors have a remarkable property.
It is well known that a two-dimensional electron gas (2DEG) can be 
easily formed on the surface of InAs and InSb.
Photoelectron spectroscopy measurements have shown that
the position of the Fermi level lies above the conduction-band minimum
at cleaved (110) surfaces with various kinds of adsorbed materials
\cite{Getzlaff2001, Betti2001}.
Recently the present authors have performed magnetotransport measurements on inversion layers formed on cleaved surfaces of $p$-type InAs \cite{Tsuji2005,Minowa2008}
and InSb \cite{Masutomi2007} covered with submonolayers of Ag or alkali metals.
The observed coverage dependence of the Hall mobility \cite{Tsuji2005,Masutomi2007} indicates that
adatoms strongly affect electron scattering in the inversion layer.
It seems feasible to probe the properties of adsorbed ultrathin films 
through transport measurements of adjacent conduction layers.

In this Letter, we report magnetotransport measurements of inversion layers
formed on {\it in-situ} cleaved InAs(110) surfaces, covered with a 
submonolayer Fe film at low temperatures.
Hysteresis in the magnetoresistance is found in a narrow coverage range.
At a coverage of 0.42~monolayer, the remanent magnetoresistance shows a clear difference between 
zero-field-cooling (ZFC) procedures and field-cooling (FC) procedures.
It also exhibits a dependence on the direction of the applied magnetic field
which corresponds to Ising-like anisotropy of the Fe film.
A long-time relaxation behavior is observed after a magnetic field sweep.
These results strongly indicate that the 2D spin-glass ordering occurs
in the submonolayer Fe film.

The InAs samples used were cut from a Zn-doped single crystal with an acceptor concentration of $1.2 \times 10^{17}$cm$^{-3}$.
Sample preparation and experimental procedures are similar to those used in Ref.~\cite{Tsuji2005}.
Cleavage of InAs, subsequent deposition of Fe and transport measurements on 
the cleaved surface (3~mm $\times$ 0.4~mm)
were performed at low temperatures in an ultrahigh vacuum chamber with a liquid ${}^4$He cryostat.
The standard four-probe lock-in technique was used at 13.8~Hz 
with two current electrodes and four voltage electrodes
prepared by deposition of gold films onto noncleaved surfaces at room temperature.
The sample was mounted on a rotatory stage to control the magnetic field direction
with respect to the surface normal.
The electron density $N_s$ and mobility $\mu$ of the 2DEG were determined from the Hall measurements
in a perpendicular magnetic field.

Figure 1(a) shows the magnetoresistance of the 2DEG
observed at $T=1.7$~K for an Fe coverage of $\Theta = 0.42$~monolayer
in a magnetic field applied parallel to the surface.
One monolayer (ML) is defined as Fe atomic density equivalent to the surface atomic density of InAs(110)
(7.75$\times 10^{14}$ atoms$/$cm$^2$),
and the absolute values of $\Theta$ in this work have uncertainties of 10~\%.
The magnetic field sweep rate was 24~mT/sec and
all the data were obtained after a waiting time of 1200~sec.
A long-time relaxation behavior will be discussed later.
A significant reduction of the resistivity $\rho$ is observed 
during the initial magnetic field cycle of 0~T $\rightarrow$ $+9$~T $\rightarrow$ 0~T.
In the subsequent cycles between $-9$~T and $+9$~T, $\rho$ follows the hysteresis loop
[see also Fig.~1(b)].
The hysteresis behavior appears only in a narrow $\Theta$ range.
The width of the hysteresis loop has a maximum at about $\Theta \approx 0.42$~ML
and becomes very small for $\Theta < 0.3$~ML and $\Theta > 0.5$~ML.
Similar $\Theta$-dependent hysteresis behavior was also observed in other samples.

\begin{figure}[t!]
\includegraphics[width=8cm]{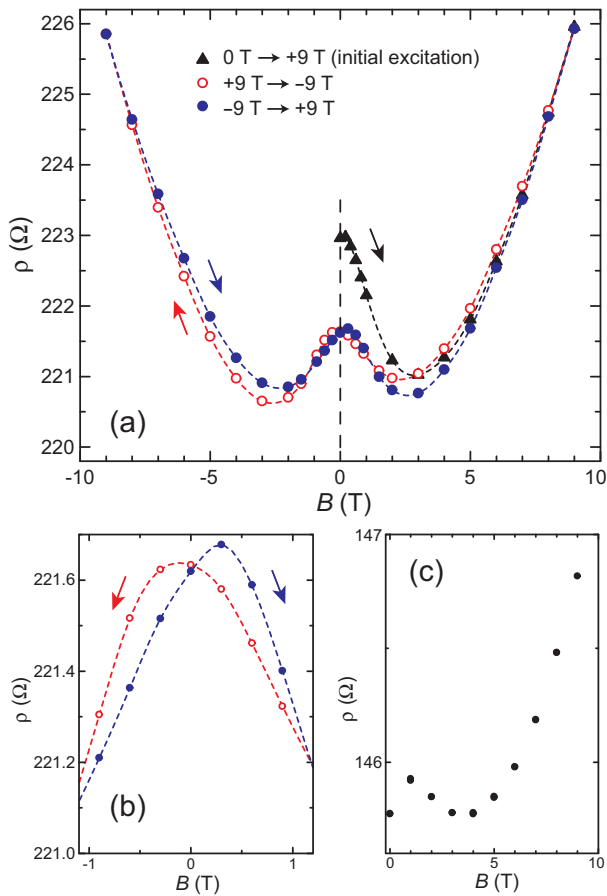}
\caption{
In-plane magnetic field dependence of $\rho$ at $T=1.7$~K.
(a)(b) Results for $\Theta=0.42$~ML.
The electron density and mobility of the 2DEG are
$N_s=3.54 \times 10^{12}~{\rm cm}^{-2}$ and
$\mu=7.9 \times 10^3~{\rm cm}^2 /{\rm V~s}$, respectively.
Solid triangles represent the data obtained during the initial excitation 
up to $+9$~T. Open (solid) circles represent the data obtained during subsequent sweeps
from $+9$~T ($-9$~T) to $-9$~T ($+9$~T).
(c) Results for $\Theta=0.17$~ML
($N_s=4.10 \times 10^{12}~{\rm cm}^{-2}$ and
$\mu=1.04 \times 10^4~{\rm cm}^2 /{\rm V~s}$).
No hysteresis was observed.
}
\end{figure}

In Fig. 1(c), we show a magnetoresistance curve at $\Theta =0.17$~ML where the hysteresis disappears.
Positive $B$-dependence in the high-$B$ region is also seen.
Since similar positive magnetoresistance appears in 2DEGs covered with nonmagnetic materials,
it should be attributed to intrinsic effects of 2DEGs,
such as the orbital effect owing to the finite thickness of the inversion layer \cite{Streda,Heisz},
or the resistivity increase induced by the spin polarization \cite{Okamoto1999,Okamoto2004}.
On the other hand, it is hard to explain the negative magnetoresistance observed
in the low-$B$ region for $\Theta=0.42$~ML [Fig.~1(a)]
in terms of the characteristics of the 2DEGs.
The $\Theta$ dependencies of $N_s$ and $\mu$ are gradual
in the range of $0.08~{\rm ML} \le \Theta \le 0.50~{\rm ML}$,
although the amplitude of the negative magnetoresistance changes drastically with $\Theta$.
We consider that the negative magnetoresistance and hysteresis are caused by
changes in the magnetic state of the Fe film.
Conduction electrons in the inversion layer move in a random potential induced 
by the spatial magnetization fluctuations of the Fe layer
unless the exchange interaction is negligible.
The negative magnetoresistance can be attributed to 
the suppression of the magnetization fluctuations
with increasing average polarization.
In contrast, an effect of the magnetization of conduction electrons on the magnetism of adsorbates is expected to be negligible
since $N_s$ is 2 orders of magnitude smaller than the atomic density of adsorbates
and the Pauli paramagnetic susceptibility is very small.

A recent calculation by Sacharow has shown that 
ferromagnetic structures are favorable in Fe$[001]$ chains
and antiferromagnetic structures are favorable in Fe$[1{\bar 1}0]$ chains
on InAs(110) \cite{Sacharow}.
Coexistence of ferromagnetic and antiferromagnetic interactions is expected
since Fe adatoms in the present sample should be randomly distributed
due to low substrate temperature deposition.
The observed hysteresis in the magnetoresistance is associated with
the irreversibility in a spin glass which appears at appropriate submonolayer coverages.
By analogy with temperature-concentration phase diagrams 
for some three-dimensional systems \cite{Youngreview},
we think that the spin-glass phase exists 
between the ferromagnetic (higher-$\theta$) and paramagnetic (lower-$\theta$) phases.
The reduction of $\rho$ during the initial magnetic field cycle cannot be related simply to the remanent magnetization
since it is not recovered by applying a reverse magnetic field.
The result suggests that a strong magnetic field has a persistent effect on the magnetization fluctuations in the spin-glass system.
The remanent magnetoresistance can be removed only by annealing the sample. 
We found that the zero-magnetic-field resistivity returns to the initial value of 223.0~$\Omega$
after a thermal cycle up to 12~K or higher.
From this we estimate a spin-glass transition temperature for $\Theta=0.42$~ML to be $T_g = 12$~K.

In general, the magnetic state of a spin-glass system depends strongly
on the external magnetic field in which the sample was cooled from $T_g$.
The difference in remanent magnetization between zero-field-cooled and field-cooled samples is observed in various materials \cite{Youngreview}.
In Fig.~2(a), the zero-magnetic-field resistivity $\rho_{0}$ at $T=1.7$~K
is shown for various cooling and magnetizing procedures.
\begin{figure}[t!]
\includegraphics[width=8cm]{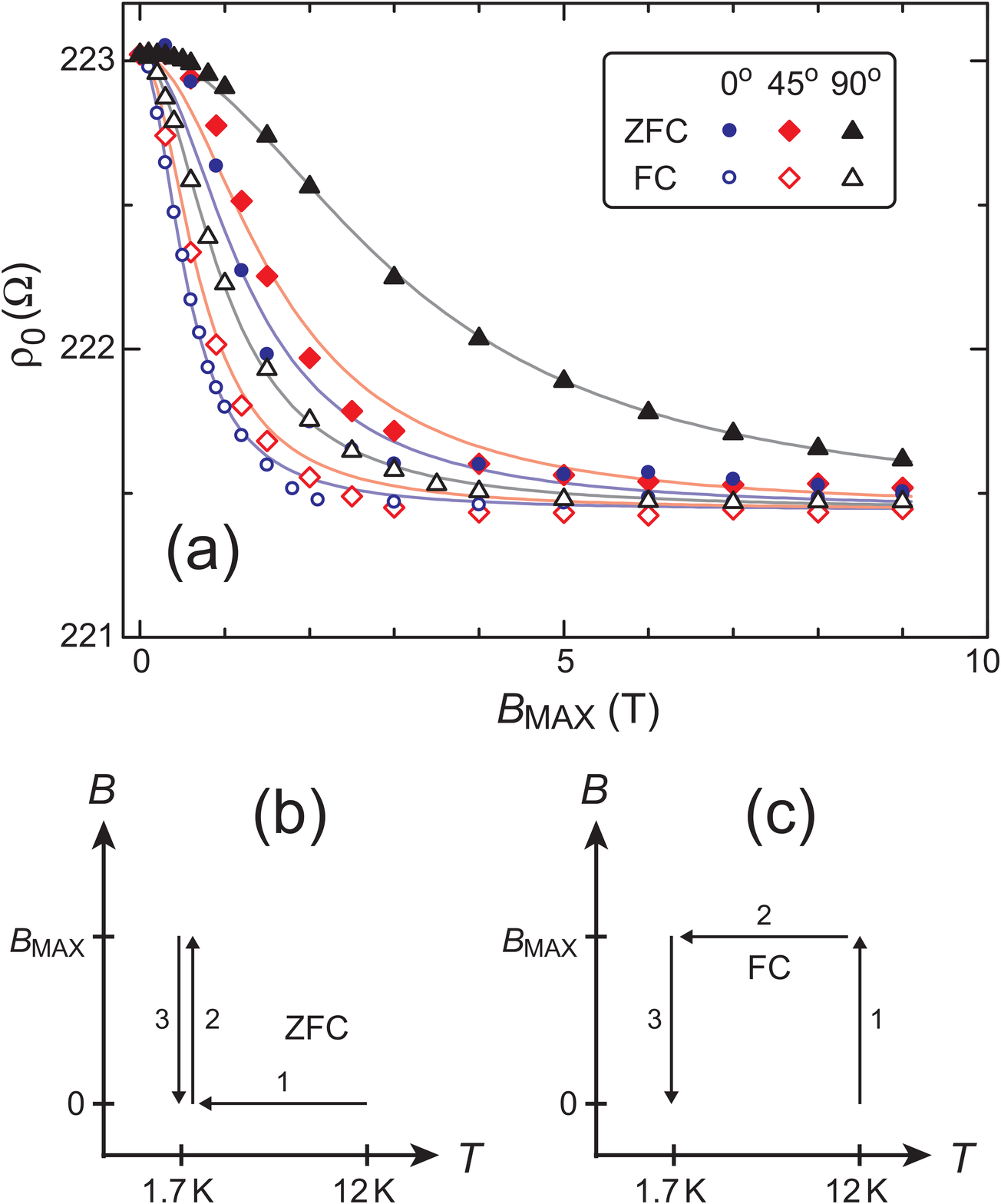}
\caption{
(a) Remanent magnetoresistivity observed at $B=0$~T and $T=1.7$~K
for different cooling and magnetizing procedures.
The filled and open symbols represent the data obtained after
ZFC procedures and FC procedures, respectively.
The magnetic field was applied at different angles of
$\varphi=0^\circ$ (circles), $45^\circ$ (diamonds) and $90^\circ$ (triangles)
with respect to the surface normal.
The data at $\varphi=0^\circ$ ($45^\circ$) are offset by $-0.10~\Omega$ ($-0.22~\Omega$)
for clarity.
Solid lines represent 
$\rho_0(B_{\rm max})=\rho_0(0)-\Delta \rho /[1+(B_0/B_{\rm max})^2]$,
where $\rho_0(0)=223.02~\Omega$, $\Delta \rho=1.58~\Omega$ and $B_0$ is a fitting parameter for each curve.
(b) ZFC procedures.
After cooling the sample, the magnetic field was applied up to a maximum value $B_{\rm max}$.
(c) FC procedures.
}
\end{figure}
In ZFC procedures,
the magnetic field excitation up to a maximum value $B_{\rm max}$
was performed at the measurement temperature of 1.7~K as illustrated in Fig.~2(b).
On the other hand, it was performed at $T_g=12$~K
in FC procedures [see Fig.~2(c)].
The cooling rate was 0.5~K/min in both procedures.
For all series, $\rho_0$ decreases as $B_{\rm max}$ increases and approaches a constant value
in a high $B_{\rm max}$ regime where the spin polarization is expected to be completed.
The observed $B_{\rm max}$-dependence of $\rho_0$ was roughly approximated by
$\rho_0(B_{\rm max})-\rho_0(0) =  [\rho_0(\infty)-\rho_0(0)] /[1+(B_0/B_{\rm max})^2]$,
where $B_0$ is a fitting parameter for each curve.
The reduction of $\rho_0(B_{\rm max})$ for FC procedures is faster than that for ZFC procedures.
This is consistent with a common feature of spin-glass systems---that
FC magnetization is larger than ZFC magnetization \cite{Youngreview}.
The reduction of $\rho_0$ depends also on the direction of the external magnetic field.
The observed anisotropy suggests that the magnetic field strength required
for the complete spin polarization is lowest
in the direction perpendicular to the surface ($\varphi=0^\circ$)
and the system has Ising-like (easy-axis) anisotropy.

A particularly interesting feature of spin glasses
is the anomalously slow relaxation.
The approximately logarithmic time dependence of the relaxation of the remanent magnetization 
has been reported for various spin-glass systems \cite{Youngreview}.
Figure 3(a) shows time evolution of $\rho$ at $B=0$
after a down sweep of the parallel magnetic field from 9~T \cite{50sec}.
Slow relaxation behavior is clearly seen \cite{otherrelax}.
\begin{figure}[t!]
\includegraphics[width=7cm]{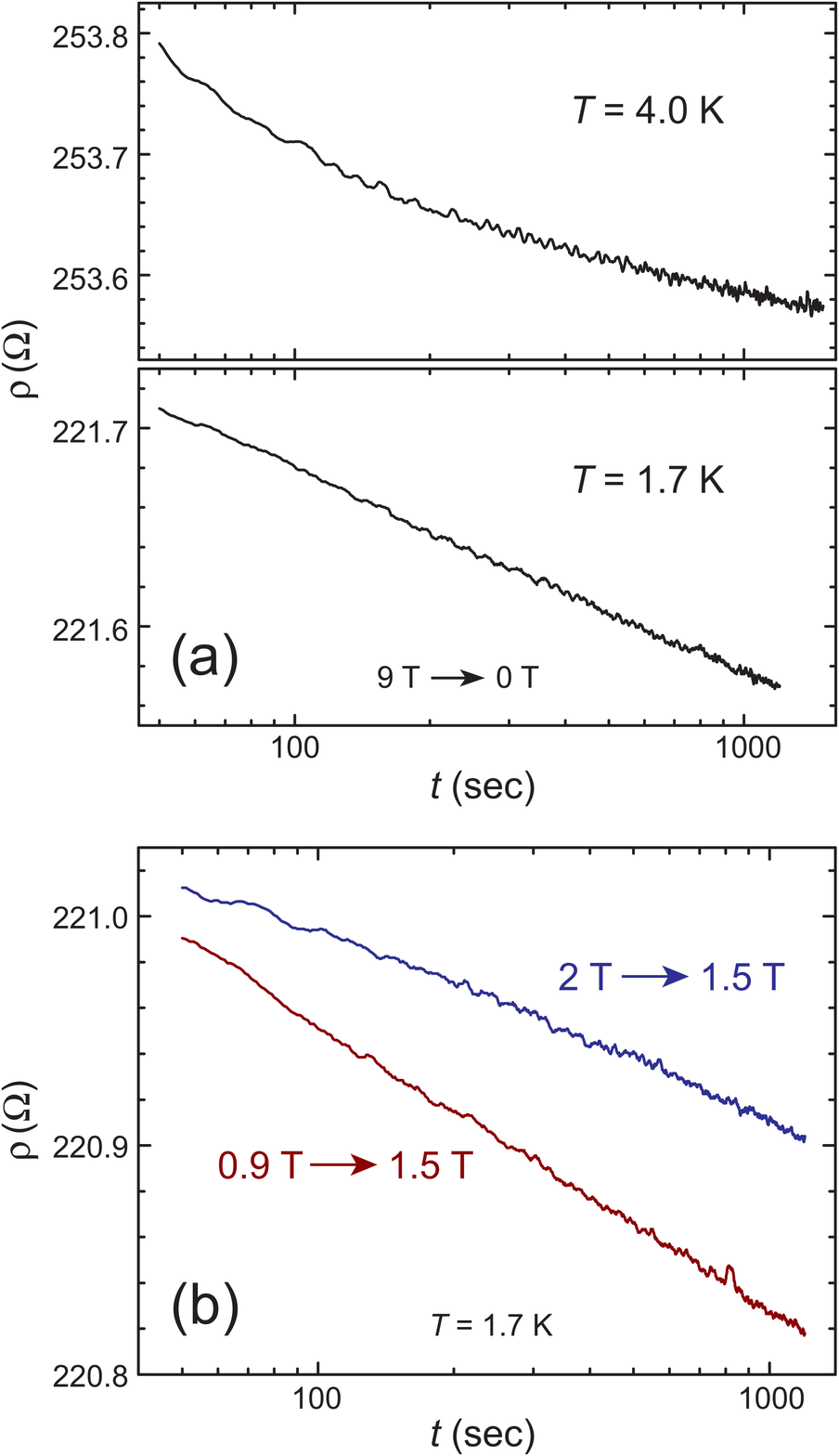}
\caption{
Time evolution of $\rho$ after magnetic field sweep at $\varphi=90^\circ$.
(a) Data obtained at 4.0~K (upper curve) and 1.7~K (lower curve) after a sweep from 9~T to 0~T.
(b) Data obtained at $T=1.7$~K and $B=1.5$~T after a down sweep from 2.0~T (upper curve)
and an up sweep from 0.9~T (lower curve).
}
\end{figure}
While the $t$ dependence is slightly curved at 4.0~K on a $\rho$ vs $\log t$ plot,
it is well fit by a straight line at a lower temperature of 1.7~K.
In Fig.~3(b), the relaxation observed at $B=1.5$~T is shown.
The sign of $d\rho/dt$ does not depend on the direction of the preceding magnetic field sweep
and is always negative.
This is in contrast to the relaxation behavior of the total magnetization \cite{Youngreview,Lundgren1983}
which can be interpreted as a delayed response to a magnetic field change.
As discussed above, the resistivity of the 2DEG is considered
as a probe of the magnetization fluctuations of the Fe layer.
The observed $t$ dependence of $\rho$ suggests that the magnetization fluctuations always decrease
when the spin-glass system relaxes toward a metastable state,
irrespective of whether the average magnetization increases or decreases.

In summary, we have studied a magnetic state of the submonolayer Fe film
through magnetotransport measurements of the 2DEG formed 
at the cleaved surface of InAs.
Hysteresis behavior is observed in a narrow coverage range around $\Theta=0.42$~ML.
It is associated with the irreversibility of a spin-glass system of Fe adatoms.
This interpretation is strongly supported by
the observations of characteristic features of spin glasses.
A clear difference between ZFC procedures and FC procedures 
is seen in the remanent magnetoresistance measurements
and relaxation after magnetic field sweep exhibits 
a logarithmic time-dependence.

We thank T. Matsui for pointing out to us Ref.~\cite{Sacharow}.
This work has partly supported by Grant-in-Aid for Scientific Research (B) (No. 18340080), Grant-in-Aid for Scientific Research on Priority Area "Physics of New Quantum Phases in Superclean Materials" (No. 20029005), and Grant-in-Aid for JSPS Foundation (No. 1811418) from MEXT, Japan.


\begin{thebibliography}{39}
\bibitem{50sec}
Data for $t < 50$~sec are not shown because the temperature of the sample stage
slightly fluctuated due to a technical problem of the temperature control system.
For $t > 50$~sec, the fluctuations were not observed
and we believe that the temperature difference
between the sample and the resistance thermometer was negligible.
After zero-field-cooling from 12~K to 1.7~K with 0.5~K/min,
time dependence was not observed at least for $t > 50$~sec 
both in the sample resistivity and the thermometer resistance.
\bibitem{otherrelax}
For $\Theta=0.33$~ML and $0.50$~ML, the $t$-dependence of $\rho$ was found to be very weak after low-temperature magnetic field sweep.
\end{thebibliography}
\end{document}